\documentclass{article}

\usepackage{arxiv}

\usepackage[utf8]{inputenc} 
\usepackage[T1]{fontenc}    
\usepackage{hyperref}       
\usepackage{url}            
\usepackage{booktabs}       
\usepackage{amsfonts}       
\usepackage{nicefrac}       
\usepackage{microtype}      
\usepackage{lipsum}
\usepackage{fancyhdr}       
\usepackage{graphicx}       
\graphicspath{{media/}}     
\usepackage{xcolor, soul}         
\sethlcolor{green}
\usepackage[normalem]{ulem}

\usepackage{censor}
\usepackage{amsmath}
\usepackage{amssymb}
\usepackage{mathtools}
\usepackage{amsthm}
\usepackage{algorithm}
\usepackage{algorithmic}
\usepackage[square,sort,comma,numbers]{natbib}
\usepackage{authblk}

\pdfoutput=1

\title{Forecasting labels under distribution-shift for machine-guided sequence design}

\author[*1]{Lauren Berk Wheelock}
\author[*]{Stephen Malina}
\author[*]{Jeffrey Gerold}
\author[*2]{Sam Sinai}
\affil[*]{Dyno Therapeutics\\
  343 Arsenal Street, Suite 101\\
  Watertown, MA, US}
\affil[1]{\texttt{lauren.wheelock@dynotx.com} }
\affil[2]{\texttt{sam.sinai@dynotx.com} }

\begin{document}
\maketitle

\StopCensoring

\begin{abstract}
  The ability to design and optimize biological sequences with specific functionalities would unlock enormous value in technology and healthcare. In recent years, machine learning-guided sequence design has progressed this goal significantly, though validating designed sequences in the lab or clinic takes many months and substantial labor.  It is therefore valuable to assess the likelihood that a designed set contains sequences of the desired quality (which often lies outside the label distribution in our training data) \emph{before} committing resources to an experiment. Forecasting, a prominent concept in many domains where feedback can be delayed (e.g. elections), has not been used or studied in the context of sequence design. Here we propose a method to guide decision-making that forecasts the performance of high-throughput libraries (e.g. containing $10^5$ unique variants) based on estimates provided by models, providing a posterior for the distribution of labels in the library. We show that our method outperforms baselines that naively use model scores to estimate library performance, which are the only tool available today for this purpose. 
\end{abstract}

\section{Introduction}
\label{sec:intro}

Biological sequence design has long been of interest to practitioners in many domains, from agriculture to therapeutics. For decades, sequences were designed through two means (i) Labor-intensive rational design where expert human knowledge would generate a handful of candidate sequences \censor{\cite{hellinga1997rational}}, (ii) High-throughput directed evolution approaches that utilize biological evolution to optimize sequences towards a desired property \censor{\cite{arnold1998design}}. Recently, the ability to synthesize DNA in high-throughput, together with the wide adoption of high-capacity of machine learning models, has opened a new path that can combine the benefits of rational design (high quality), and directed evolution (high throughput) \censor{\cite{wittmann2021advances, yang2019machine,bryant2021deep, alley2019unified, hsu2021combining}}. In this setting, libraries containing up to $10^5$ sequences are designed using machine learning algorithms. Machine learning methods are used to score, optimize, and filter sequences before committing to experiments \censor{\cite{chowdhury2022single,shin2021protein,bryant2021deep, brookes2019conditioning, sinai2021generative}}. In recent years, increasingly nuanced perspectives on how to improve our trust in the output of machine learning models and paired optimization procedures have evolved \censor{\cite{fox2005directed, romero2013navigating, brookes2019conditioning,kumar2020model,  angermueller2020population, fannjiang2022conformal, nijkamp2022progen2,feng2022designing}}. 
Using these methods, sequences targeting different objectives can be synthesized (e.g. transcription factor binding or other regulatory sequences \censor{\cite{barrera2016survey,vaishnav2022evolution}}) in a library that can be measured in the desired context. However, especially with products or traits of high complexity (e.g. \emph{in-vivo} studies of proteins \censor{\cite{ogden2019comprehensive}}), the overall cost required to validate designs can be prohibitive. 

Therefore, even with model evaluations and calibration of uncertainty around samples, there remains a gap in our ability to \emph{forecast} the probability of success: be it reaching a certain maximum performance, or finding a certain number of variants above a minimum desired performance. This is distinct from attempting to predict the performance of single sequences, in that it focuses on predicting the right-tail distribution of the performance our entire library. In many settings, we would want to know whether the experiment has a high chance of finding a (generally rare) high-performing sequence overall. Forecasts can help us decide whether to commit to a certain design and can save large costs. Forecasts can also inform other decisions such as deciding whether to repeat the design procedure for a library, deciding among libraries designed for different targets, or estimating the final price of developing a drug. 

Forecasting is ubiquitous in domains with delayed feedback such as elections \censor{\cite{hummel2013fundamental, wang2015forecasting}}. The related topic of label shift \censor{\cite{lipton2018detecting, saerens2002adjusting}} classically relies on the ``anticausal'' assumption that the distribution of inputs given labels is constant across training and test sets - an assumption that is invalid in the case of design.  More generally, domain adaptation has been studied in biological sequence design \censor{\cite{abestesh2021efficient}} but does not directly address forecasting and calibrating distributions under covariate and label shift. Recent work in conformal prediction directly tackles the problem of the kind of covariate shift that arises in sequence design settings \censor{\cite{fannjiang2022conformal}}, but its use requires known probability distributions over training and test sets, and nonzero prior probabilities on the entire support, meaning it cannot be applied to most libraries that were not designed with this approach in mind.  To our knowledge, there are currently no methods that are suitable for forecasting library performance in the sequence design setting. This setting presents an interesting and somewhat unique challenge. For every designed sequence we can obtain scores for the expected performance, possibly from multiple models. However we are often aiming to make sequences that have a significantly higher score than anything observed in our training data, i.e.  distribution shift is by design. Our challenge is to find the right balance between trusting our models' predictions out-of-distribution and betting that our new designs would provide us with better-than-observed sequences.

\section{Forecasting method overview}
\label{sec:method}

We start with labeled training data $(\mathbf{S}^0, \mathbf{Y}^0)$, where $\mathbf{S}^0=\{s_i^0\}$ is a set of biological sequences and $\mathbf{Y}^0 = \{y_i^0\}$ is a set of continuous-valued labels, generally a fitness measurement in the sequence design setting such as packaging or transduction efficiency rates.  Our goal is to forecast a distribution of labels $\mathbf{Y}^1$ for an unlabeled set $\mathbf{S}^1$. That is, we are not concerned with the accuracy of each pair $(s_i^1, y_i^1)$, but only the overall distribution of $\mathbf{Y}^1$, and in particular, in the right tail of $\mathbf{Y}^1$, which indicates the maximum quality achieved by the set of sequences. To create our forecast, we have at our disposal a set of $J$ regression models trained on $(\mathbf{S}^0, \mathbf{Y}^0)$, which produce test set predictions $m_{ij}$ for sequence $i$ by model $j$.  

Naively, we could form ensembled point estimates $\hat{y}_i = \sum_j m_{ij}/J$ for each sequence and predict the distribution of $\mathbf{Y}^1$ to be the distribution of $\hat{y}_i$.  There are three main disadvantages to this approach, which inspire different aspects of our forecasting method.  We address them briefly below, and give a more thorough treatment with a complete algorithm in Section \ref{sec:method-details}.

\textbf{1) An implicit unimodal Gaussian assumption} Empirically, model ensembles tend towards unimodal Gaussian score distributions which do not empirically fit experimental data from designed sequences well.  At several points in the experimental pipeline variants may ``drop out,'' failing to produce enough signal to reliably approximate a label (for example, due to failure of a protein to fold).  This results in a multimodal distribution at both the population level, and implicitly, at the level of each sequence's posterior.  Thus, we seek to model each sequence as a bimodal Gaussian Mixture Model (GMM), and learn the parameters for each sequence's posterior from its model scores.  Explicitly, given an i.i.d. set of model predictions $m_{ij}$ for sequence $i$, we seek to infer a probability $p_i$ that distinguishes between the two distributions in the Gaussian mixture, as well as a mean and variance parameter for each Gaussian mode.  Moreover, while we use a GMM, our method could in theory be applied using a range of more complex distributions, with the only constraint being our ability to sample from them.

\textbf{2) Distribution (covariate and label) shift} Typically, the sequence set $\mathbf{S}^1$ is designed with model-guided exploration strategies informed by $(\mathbf{S}^0, \mathbf{Y}^0)$, with the objective of producing sequences that outperform the best sequences in $\mathbf{S}^0$.  This results in both significant distribution (covariate) shift, because the sequences $\mathbf{S}^1$ are reaching into untested areas of sequence space, and label shift, since we anticipate that $\mathbf{Y}^1$ will dominate  $\mathbf{Y}^0$, both on average and among each sets top-performers.  (While there has been some important recent work on prediction in this design setting \citep{fannjiang2022conformal}, this work assumes a shift in distribution within a consistent domain between $\mathbf{S}^0$ and $\mathbf{S}^1$ and no label shift.)  To address distribution and label shift, we start by applying non-parametric non-linear calibration techniques to produce a ``conservative'' forecast that still allows for some label shift due to model uncertainty.  We then consider scenarios with some trust placed in raw model scores to allow for some amount of extrapolation to regions further from our training set.  

\textbf{3) Point estimates to posteriors} The point estimates that arise from model ensembles do not provide a posterior for $\mathbf{Y}^1$ (nor do the model score variances, directly), and consequently these tend to underestimate the frequency of events that are rare at the sequence level, but common at the population level, such as the occurrence of high-valued sequences in the library.  In our method, we simulate draws of the entire library from the sequence-level posteriors to produce both expected distributions as well as the frequency of rare events, which we interpret as posterior probabilities.

\section{Forecasting method description \label{sec:method-details}}

To generate forecasts, we first transform model predictions for each sequence into parameters for their posterior distributions.   We then draw from those sequence-level posterior distributions to form simulations of the library, generating a library-level posterior.  This library-level posterior reflects our epistemic uncertainty about the ground-truth performance of each sequence given our model predictions and the aleatoric uncertainty of our measurements given this ground-truth.  That is, our predictions are more like prediction intervals than confidence intervals, and we do not generate a posterior for ground-truth values.  Our objective is to infer sequence-level posterior parameters from model predictions in a way that is both well-calibrated to the training data and allows for test set performance of the sequences to differ from or exceed training set performance due to distribution shift.

\subsection{Fitting sequence label posteriors}

While this forecasting framework can be applied to learn posteriors for many distributions families, informed by empirical library label distributions from historical experiments, we believe the natural distribution for sequence labels in our data is a Gaussian mixture model (GMM) with two modes: one for ``functional'' sequences and one for ``non-functional'' or ``broken'' sequences.  (To arrive at this conclusion, we considered a number of alternative distribution families, including varying the skew and kurtosis of each mode of the GMM, but did not find sufficient improvements to justify the increased model complexity.)

To model a sequence $s_i$ using a GMM, we assume there is a probability of functionality $p_i$, a mean and variance in the functional mode $\left(\mu_i^+, (\sigma_i^+)^2\right)$, and a mean and variance in the non-functional mode $\left(\mu_i^-, (\sigma_i^-)^2\right)$ that parameterize normal distributions $\mathcal{N}$ so that
\begin{equation} \label{eq:bimodal}
    Y_i \sim p_i \mathcal{N}(\mu_i^+, \sigma_i^{+}) + (1-p_i) \mathcal{N}(\mu_i^-, \sigma_i^{-}).
\end{equation}

In contrast, our predictive models only provide point estimates $m_{ij}$ for each sequence.  We assume that the true, multimodal distribution of each sequence can be summarized with two degrees of freedom (a mean $\mu_i$ and standard deviation $\sigma_i$) and that these two parameters independently generate model scores, mixture model parameters, and measurement values.  Explicitly, we assume the model predictions $m_{ij} \sim \mathcal{N}(\mu_i, \sigma_i)$, so that, given a set of model predictions, we infer $\mu_i$ and $\sigma_i$ to be the models' sample mean and variance ($\mu_i = 1/J \sum_j m_{ij}$ and $\sigma_i^2 = 1/J \sum_j (m_{ij}-\mu_i)^2 $). We further assume there are independent relationships between $\mu_i$ and the set ($p_i$, $\mu_i^+$, $\mu_i^-$), and between $\sigma_i$ and the pair $\sigma_i^+$ and $\sigma_i^-$. 

Since the GMM parameters $(p_i, \mu_i^+, \mu_i^-, \sigma_i^+, \sigma_i^- )$ are unique to each sequence, we cannot infer them in the usual manner using $\mathbf{S}^0$ as a training set.  Instead, we need to further model and learn the relationship between the pair $\mu_i, \sigma_i$ and the GMM parameters.  Specifically, we start by identifying the value $y_{mid}$ that we use to separate the two modes of $\mathbf{Y}$.  This value can either be set manually, using expert knowledge, or automatically by analyzing the distribution $\mathbf{Y}^0$.  We found that Otsu's method \censor{\cite{otsu1979threshold}}, which finds the separating point that minimizes intra-class variance, provided robust values of $y_{mid}$ on our data. We can then divide our set $\mathbf{S}^0$ into two halves across the boundary: $\mathbf{S}^{0+} = \{s_i \mid y^0_i \geq y_{mid} \}$ and  $\mathbf{S}^{0-} = \{s_i \mid y^0_i < y_{mid} \}$.  This provides us with separate training sets for the functional parameters $\mu_i^+, \sigma_i^+$ and the non-functional parameters $\mu_i^-, \sigma_i^-$.

\subsection{About isotonic regression}

We will run isotonic regression to find the best monotonic piece-wise linear fit to this data.  Explicitly, isotonic regression operates on a dataset of pairs of scalars $\{x_i, y_i\}$ and produces a non-parametric model represented by data-prediction pairs $(x_i, \hat{y}_i)$ that seek to minimize the least squares error $\sum_i (\hat{y}_i-y_i)^2$ subject to a monotonicity constraint $y_i \leq y_{j}~\forall i,j~\text{s.t.}~x_i<x_j$.  This results in a quadratic program, though it is easily solvable exactly by sorting $y_i$ and iteratively averaging pairs of ``violators'' of monotonicity, making training efficient and deterministic \citep{friedman1984monotone}.

As a point of notation, we will use the abbreviation $IR_y$ to refer to an isotonic regression model trained to predict $y$ defined by the pairs $(x_i, \hat{y}_i)$, and $IR_y(x_i')$ to be the prediction of this model given input $x_j'$.  To compute $IR_y(x_j')$ on a new data point, first we check to see if $x_j'=x_i$ for some $i$ and if so we predict the corresponding $IR_y(x_j') = \hat{y}_i$.  Otherwise we sort $x_i$ and find a consecutive pair such that $x_i < x_j' < x_{i+1}$ and predict $IR_y(x_j')$ by linearly interpolating between $\hat{y}_i$ and $\hat{y}_{i+1}$.  If $x_j'$ is less than all $x_i$, or more than all $x_i$, $IR_y(x_j')$ is set to the min and max values of $\hat{y}_i$ respectively.  We note here that this necessarily means that isotonic regression will never produce labels outside the range of the training labels - we will address this in a few ways in the coming sections.

\subsection{Inferring parameters with isotonic regression}

We assume a non-linear monotonic relationship between the model ensemble mean for a sequence $\mu_i = 1/J \sum_j \mu_{ij}$ and the probability that the sequence $i$ will be functional ($p_i$).  To infer $p_i$, we train an isotonic regression model $IR_p$ that, given $\mu_i$, aims to predict the indicator $I_i^+$ which is 1 if $s_i>y_{mid}$ and 0 otherwise.  Effectively, given a new input $\mu_i$, this model returns the fraction of sequences that are functional out of the training samples with similar mean values, and interprets this rate as the probability that the sequence $i$ will be functional.

Inferring the mean parameters $\mu_i^+$ and $\mu_i^-$ is more straight-forward: we build isotonic regression models $IR_{\mu^+}$ and $IR_{\mu^-}$ to predict $y_i$ from $\mu_i$, but restrict the training set to $\mathbf{S}^{0+}$ and $\mathbf{S}^{0-}$ respectively. This gives us calibration to the conditional distributions for being functional and non-functional respectively.

To infer the variances $\sigma_i^{2+}$ and $\sigma_i^{2-}$, we first form squared residuals of labels given model ensemble means $res_i = (y_i-\mu_i)^2$ and build isotonic models $IR_{\sigma^+}$ and $IR_{\sigma^-}$ relating the model variance to these residuals $res_i$.  As with $\mu_i^+$ and $\mu_i^-$, we compute $\sigma_i^{2+}$ and $\sigma_i^{2-}$ by training models on the disjoint training sets $\mathbf{S}^{0+}$ and $\mathbf{S}^{0-}$ respectively.  The complete algorithm for inferring the GMM parameters is described in Algorithm \ref{alg:theta_inference}.

\paragraph{Applying the forecast to non-ensembles} While the presentation of our method assumes access to an ensemble of models, we note that thus far the only information we have used from the ensembles is the ensemble mean and variance ($\mu_i, \sigma_i^2$) for each feature.  Therefore, as an alternative, any single model that itself outputs an expected value and uncertainty (which includes many neural networks) can stand alone in providing the input ($\mu_i, \sigma_i^2$) to forecasting calibration.  The only technique that does not generalize from ensembles to models-with-uncertainty is the ``optimistic model de-ensembling'' technique discussed in the Section \ref{sec:tuning}.

\subsection{Simulating the posterior distribution \label{sec:extensions}}

Given the parameters generated by Algorithm \ref{alg:theta_inference}, we can draw samples $\hat{y}^1_i$ for each sequence, and aggregate them into draws for the entire distribution $\hat{\mathbf{Y}}^1$.  We can then treat the set of simulated values of $\hat{\mathbf{Y}}^1$ as a posterior distribution and query this distribution to determine the frequency of distribution-level events.  By computing metrics on  $\hat{\mathbf{Y}}^1$ and considering their distributions across simulations, we can arrive at empirical confidence intervals for metrics such as the count of sequences that perform above some threshold value, as we see in Figures 1b,d,f,h.

\subsection{Tuning the forecast from conservative to optimistic \label{sec:tuning}}

We can further refine this basic algorithm using additional techniques that allow us to diversify our approach over degrees of trust in our training set.

\paragraph{Semi-calibrated regression} 
Our main calibration tool, isotonic regression, aggressively limits predicted labels to be within the range of training values.  To allow for some distribution shift, we can gradually transition from calibrated predictions towards the center of the distribution of $\mathbf{S}^1$ towards uncalibrated, out-of-distribution values towards the limits of the distribution, in a technique we call ``semi-calibration.''

Let $P_Y(y)$ be the percentile of the value $y$ from among the empirical distribution $Y$.  That is, $P(y)$ is the fraction of $y \in Y$ with $y_i < y$.  In our case, we consider the distribution of model ensemble means on our training set $\mathbf{S}^0$, that is, the set $M = \{1/J \sum_j m_{ij} \mid s_i \in \mathbf{S}^0\}$.  Then given a new sequence $s_i$ we can compute its model ensemble prediction $\mu_i$ as well as its functional isotonically calibrated mean $\mu_i^+$, and evaluate where its model ensemble falls relative to the training distribution by computing the percentile $P_M(\mu_i)$.  Finally, for any temperature-like coefficient $0 <q\leq1$, we define our semi-calibrated mean $\tilde{\mu}_i(q)$ to be
\begin{equation} \label{eq:semicalibrated-mean}
    \tilde{\mu}_i(q) = (qP_M(\mu_i))( \mu_i) + \left(1-qP_M(\mu_i)\right) (IR_{\mu_i^{+}}(\mu_i)).
\end{equation}

Thus, lower values will be completely calibrated to the training set, while higher values will be a mix of calibrated and uncalibrated values.  Note that we only produce this correction for functional mean values $\mu_i^+$, as we expect non-functional values to be fully in the training set distribution.  This leads to an update to the model from Equation \ref{eq:bimodal}:
\begin{equation} \label{eq:bimodal2}
    \hat{y}_i \sim p_i \mathcal{N}(\tilde{\mu}_i(q), \sigma_i^{+}) + (1-p_i) \mathcal{N}(\mu_i^-, \sigma_i^{-}).
\end{equation}

\paragraph{Correcting for covariate shift} In addition to model score distribution shift, we also see covariate shift that creates model score bias.  In our context, we consider edit distance to wild type the primary such covariate, though the method easily generalizes to more complex distance metrics (such as BLOSUM \censor{\cite{henikoff1993performance}}), as well as other quantitative side-information.
To correct for this shift, we form \emph{signed} residuals from the training set between the \emph{calibrated} $\mu_i^+$ values and the true values $y_i$ (i.e. $(\mu_i^+-y_i)$.  (If we are also applying the semi-calibration technique from the last paragraph, we use $(\tilde{\mu}_i-y_i)$ instead.)  
We can regress those residuals on edit distance ($ED_i$) using either isotonic or linear regression, and apply this correction back to the mean prediction $\mu_i^+$.  We can also apply this approach to adjust the probability parameter $p_i$, encoding the understanding that sequences are less likely to be functional at higher distances from the wild-type.  That is, we compute:
\begin{align} 
\begin{split}
    res_i &= \mu_i^+-y_i \\
    IR^{ED} &= \text{ IR model trained on } \{(ED_i, res_i)\} \\
    \mu_i^{ED} &= \mu_i^+-IR^{ED}(ED_i) \\
    \hat{y}_i &\sim p_i \mathcal{N}(\mu_i^{ED}, \sigma_i^{+}) + (1-p_i) \mathcal{N}(\mu_i^-, \sigma_i^{-}).
\end{split}
\end{align}

\paragraph{Optimistic model de-ensembling} So far, we have assumed model scores $m_{ij}$ are drawn from a Gaussian distributions parameterized by $\mu_i, \sigma_i$.  Alternatively, we could assume that each model $j$ represents a distinct distribution, and that $\mu_i$ are drawn from these distributions with equal probability.  Using this approach, in each simulation we first randomly select one model independently for each sequence and use that model's prediction as the sequence's expected value: $\mu_i \ m_{ij}$.  This can result in more optimistic forecasts when scores have high inter-model variance.

\paragraph{Hedging against calibration assumptions} Together, these calibration techniques create a menu of options that allow us to build forecasts that range from conservative to optimistic given input data.  Given a set of calibration strategies, we can simulate instances of $\mathbf{Y}^1$,
and by aggregating simulations across frameworks, we can form a posterior for the distribution of $\mathbf{Y}^1$ that captures our uncertainties at the sequence level, the model level, and the overall forecasting approach level.

\section{Experimental results}
\label{sec:results}

We validate our method by conducting experiments on four datasets - a set of simulated RNA binding landscapes that allow us to access ground truth values for every sequence (and repeat multiple experiments) as well as three experimentally-measured assays of protein fitness landscapes. These are a viral protein packaging landscape, an experimentally measured IgG-Fc binding dataset for Protein G's GB1 domain, and an experimentally measured GFP fluorescence dataset. Of the experimentally measured landscapes, the viral protein assay is precisely conducted in the manner that forecast is intended to be used. The GFP and GB1 landscapes are not conducted in this way, but we still expect (and observed) forecast to outperform model point estimates.

\subsection{Descriptions of experimental data \label{sec:data-details}}

\subsubsection{Simulated RNA landscapes \label{sec:landandstarts}} 

Our first set of experiments investigates the performance of forecasting approach using FLEXS \censor{\cite{sinai2020adalead}}, a simulation environment for sequence design which gives access to ground-truth and model-approximated fitness landscapes. We study design problems on two RNA landscapes with a hidden binding target of size 14 and 50 nucleotides. Each training set $\mathbf{S}^0$ was constructed by mutating a sequence from a starting seed (5 seeds for each landscape) with between 1 and 3 mutations per sequence on average.  We trained four kinds of predictive models on one-hot encoded sequences: linear regressions, convolutional neural networks, multi-layer perceptions, and random forest models.  We used four exploration algorithms to design sequences using these models $\mathbf{S}^1$: CMA-ES \censor{\cite{hansen2001completely}}, CbAS \censor{\cite{brookes2019conditioning}},Adalead \censor{\cite{sinai2020adalead}}, and random sampling.

\subsubsection{In-vitro AAV packaging assay}

Bryant et al. \censor{\citep{bryant2021deep}} quantitatively assay the packaging efficiency of 200K viral capsid variants, modified in a 28 amino-acid region of the protein. The experiment was designed in two steps, where first a smaller set of training examples were assayed and used to train \emph{classification} models. Then these models were used to design a set of variants that were optimized for the probability of packaging. The paper uses classification models and greedy optimization to generate the second batch. The existence of this first training set, and the distribution-shifted designed set is exactly the setting we have devised the our forecasting method for, and where its performance can be most meaningfully evaluated. 

For this data set we retrain \emph{regression} models on the first set of sequences ($\mathbf{S}^0$) using five independently seeded convolutional neural networks, and five recurrent neural networks. We used these models to generate ensembled point estimates as well as forecast the distribution of packaging efficiencies on the model-designed set of sequences ($\mathbf{S}^1$).

\subsubsection{Protein G GB1 IgG-Fc binding domain}

A region of four amino acids in Protein G (GB1) is known to be critical for IgG-Fc binding and has been  used extensively as a tool for evaluating sequence landscape prediction and design tasks \censor{\citep{smith2011predicting, kmiecik2008folding}}.  Our data is sourced from experiments by Wu et al. \censor{\citep{wu2016adaptation}}.  We created $\mathbf{S}^0$ by selecting sequences with performance below that of the wild-type, combined with a small fraction of sequences between wild-type and the median performance in the set, leaving all other sequences for $\mathbf{S}^1$. We trained five independently seeded random forest models and five multi-layer perceptrons on $\mathbf{S}^0$ (forgoing more sophisticated models due to the short length of the variable sequence), and used these to generate point estimates and forecast predictions for $\mathbf{S}^1$.

\subsubsection{Green Fluorescent Protein}

The green fluorescent protein of Aequorea victoria (avGFP) provides an additional fitness landscape for studying sequence prediction and design. We used a dataset of 540,250 protein variants and their associated fluorescence level \censor{\citep{sarkisyan2016local}}.  We followed the same procedure as GB1 for splitting $\mathbf{S}^0$ and $\mathbf{S}^1$, putting sequences with fluorescence below 3x log-fluorescence of the wild-type and a small portion of sequences up to the median fluorescence above this threshold into $\mathbf{S}^0$, and the rest into $\mathbf{S}^1$.  We then trained the same models as in our AAV experiments - five independently seeded convolutional neural networks, and five recurrent neural networks, and generated point estimates and forecasts for $\mathbf{S}^1$ using these models.

\begin{figure*}[!ht]
  \includegraphics[width=\textwidth]{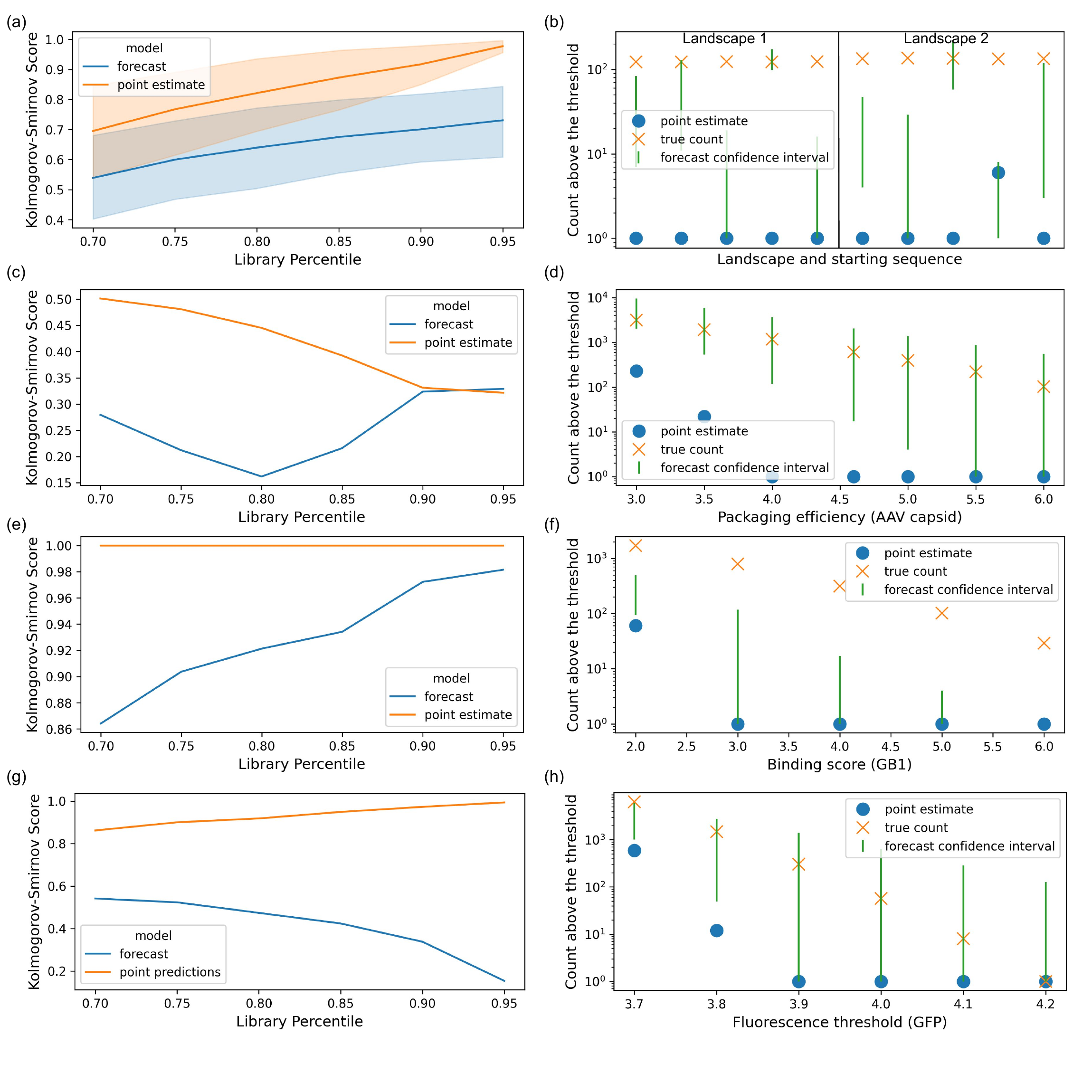}
  \caption{Comparison between forecast and point predictions in describing the right tail statistics of designed libraries. a) Distance from true (right-tail) distribution as measured by KS two-sample score for ensemble point estimate and our forecast on two RNA landscapes and five distinct designs per landscape (10 total forecasts) b) Top centile confidence interval coverage for RNA landscape 1 (14 nt) and RNA landscape 2 (50 nt) c,e,g) Distance from true distribution fit for ensemble point estimate and forecast d,f,h) Confidence interval coverage based on the number of samples above a certain measured performance c,d) For the AAV capsid design problem e,f) For the GB1 binding landscape g,h) For the GFP florescence landscape.}
  \label{fig:megafigure}
\end{figure*}

\subsection{Experimental design}

For each of these experiments, we generated training and test sets $\mathbf{S}^0, \mathbf{Y}^0$ and $\mathbf{S}^1, \mathbf{Y}^1$, and models $\mathbf{M}$.  We used the training set, models, and unlabeled test data to generate forecasts, and evaluated them against realized distributions of $\mathbf{Y}^1$ using two key tools: a 2-sample Kolmogorov-Smirnov statistic measuring distribution fit of top percentiles, and confidence interval coverage for counts of points measured above a fixed threshold value.  Both evaluate fit in the right tail of the distribution, which is both the most challenging region to predict, and the one that is most critical to sequence design applications.  We present results of forecasts on all four datasets in Figure 1, with one dataset per row of figures (RNA, AAV, GB1, and GFP respectively). While these experiments demonstrate the efficacy of the forecasting procedure in a variety of experimental settings, we acknowledge that additional experiments in follow-on studies could help clarify the method's strengths and weaknesses.  

\subsection{Discussion of results}

\begin{figure*}[!ht]
\begin{centering}
  \includegraphics[width=.8\textwidth]{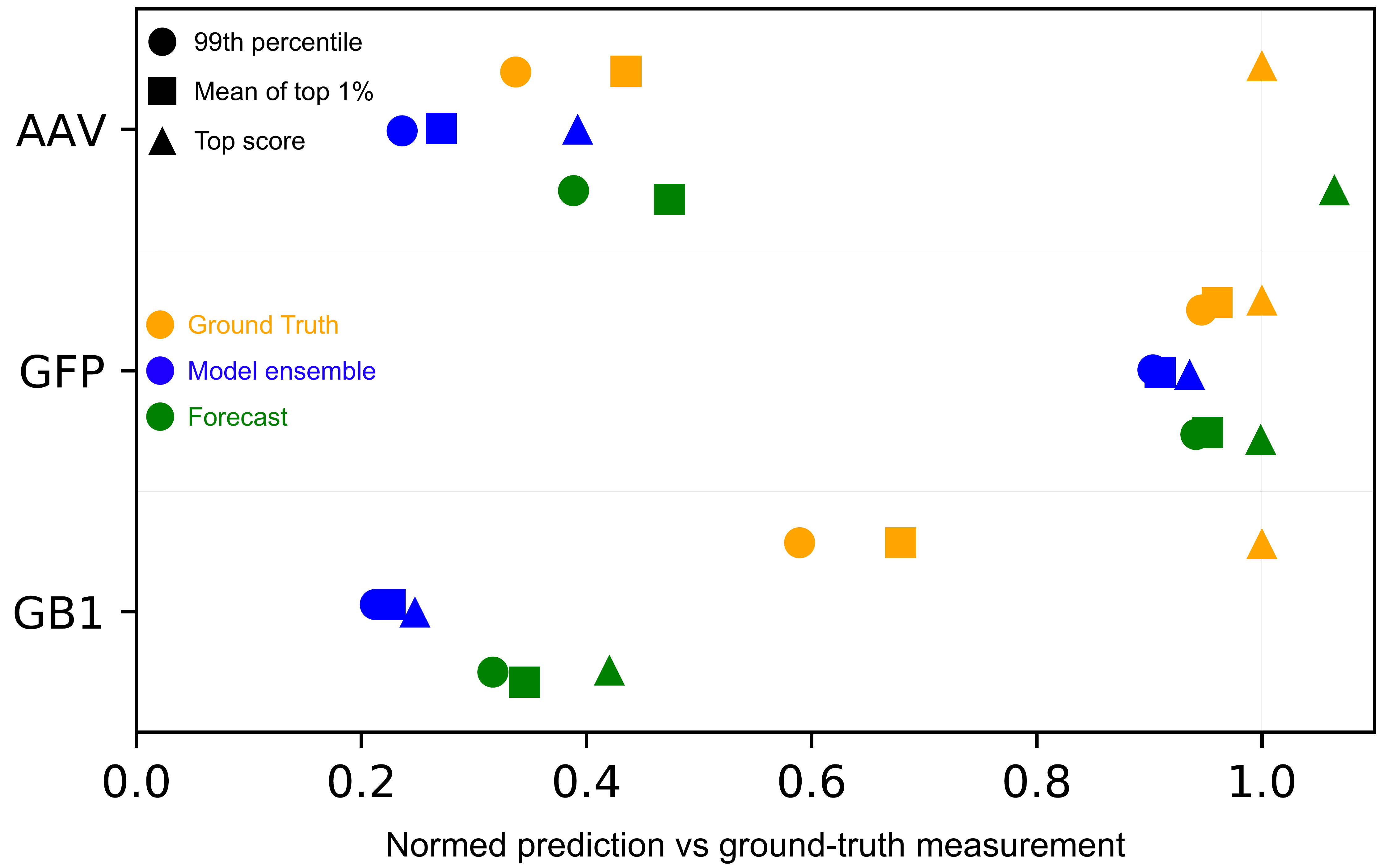}
  \caption{The ensemble, forecast, and measured values for the 99th percentile, mean of the top percentile, and maximum value for the AAV, GFP, and GB1 experiments, normalized to the maximum ground-truth measured value for each experiment.}
  \label{fig:main_agg_plot}
\end{centering}
\end{figure*}

In the left column of plots in Figure \ref{fig:megafigure} we report on the Kolmogorov-Smirnov fit between the true distribution of scores $\mathbf{Y}^1$ and either the forecast or the set of point estimates $\mu_i$ from model ensemble scores.  In figure 1a we plot the mean and 95\% confidence interval across landscapes and starts (see Sec \ref{sec:landandstarts} for details), while the other experiments have a single landscape each.  Since we are primarily concerned with fit in the right tail, we limited each distribution to values above a percentile threshold, and varied that threshold between 50\% and 95\%.  Across this range, the forecasting method improved distributional fit compared to ensemble point estimates, and while point estimates typically decayed towards 1.0 (the theoretical worst-case upper bound of the statistic), the forecast consistently maintained some predictive power even in the top 5th percentile.  A sharp covariate shift in the AAV capsid design sets $\mathbf{S}^0$ and $\mathbf{S}^1$ account for the problem difficulty in figures 1e and 1f, though even in this problem our method directionally improves upon model estimates.

In the right column of figures we focus more closely on the right tail of the label distribution of $\mathbf{Y}^1$, reporting the forecast's confidence interval for the number of sequence we can expect to find above a threshold value compared to the estimate from model ensembles and the true counts.  Since Figure 1b encompasses several landscapes and seeds, we set one threshold per landscape/seed at the 99th percentile, while in the remaining experiments with a single landscape we evaluate accuracy over a range of thresholds.  Here we see that the forecast gives confidence intervals that include the true count of sequences above a high threshold some of the time, and always improves upon the ensemble estimates, which for most datasets severely underpredicts the prevalence of top-performers.

A key item of interest for sequence design is the performance of the top variants. We report ensemble, forecast, and true values for the 99th percentile, mean of the top percentile, and maximum value for the AAV, GB1, and GFP experiments in Figure \ref{fig:main_agg_plot}.  Results from each individual RNA experiment can be found in Figure \ref{fig:rna_agg_plot} in the Appendix.  These results echo the conclusions from Figure \ref{fig:megafigure}, showing highly accurate predictions on the AAV and GFP landscapes, and directionally correct adjustments on GB1 and RNA landscapes.

\section{Conclusion}

In this paper we argued for the relevance and impact of forecasting in the sequence design setting. We developed a novel approach for forecasting label distributions under covariate and label shift that occurs during model-guided design. Our approach can be used on any machine-guided library design for which we have regression models.  We applied these methods in simulated and real-world sequence design settings and showed near-universal improvement (and never worse than the naive approach) in our ability to predict the shape of the right tail and counts of top performers. This work enables valuable estimation of the quality of designed libraries of biological sequences long before experimental results can be produced, which provides essential feedback to the designer.  We hope that by defining this problem framework, and showcasing an approach to address it, we inspire further development for improving distributional forecasting in the model-guided sequence design setting.

\bibliography{main.bib}
\bibliographystyle{unsrt}

\appendix
\section{Appendix}

\subsection{Computational details}

\subsubsection{Compute}
All of our experiments were run using a single server with a single GPU running in GCP (Google Cloud Platform). We used an Nvidia V100 for training models on the GFP landscape and an Nvidia K80 for the other three experiments' model training.

\subsubsection{Hyperparameters}
Across all of our experiments, we used five model architectures: convolutional neural networks (CNNs), recurrent neural networks (RNNs), multi-layer perceptrons (MLPs), linear models, and random forests. Linear models and random forests were initialized with default parameters using the sklearn library. CNNs used 32 filters, 64 filters, and 256 filters with 1, 2, and 2 convolutional layers followed by 1, 2, and 2 hidden layers of width 32, 64, and 64 for the AAV, RNA, and GFP experiments respectively. RNNs used embeddings of size 32 combined with 1 and 2 recurrent layers, then followed by 1 hidden layer of size 56 and 128 for AAV and GFP respectively. MLPs used 1 and 3 hidden layers of width 50 and 32 for GB1 and RNA experiments respectively. All three model architectures were trained used Adam with a learning rate of 1e-3 across experiments.

\subsubsection{Licenses}
FLEXS is open source and Apache licensed.  All other code was written for this project in python using common packages that use BSD, PSFL, Apache, and MIT licenses.

\subsection{Additional RNA results}

\begin{figure*}[!ht]
\begin{centering}
  \includegraphics[width=.8\textwidth]{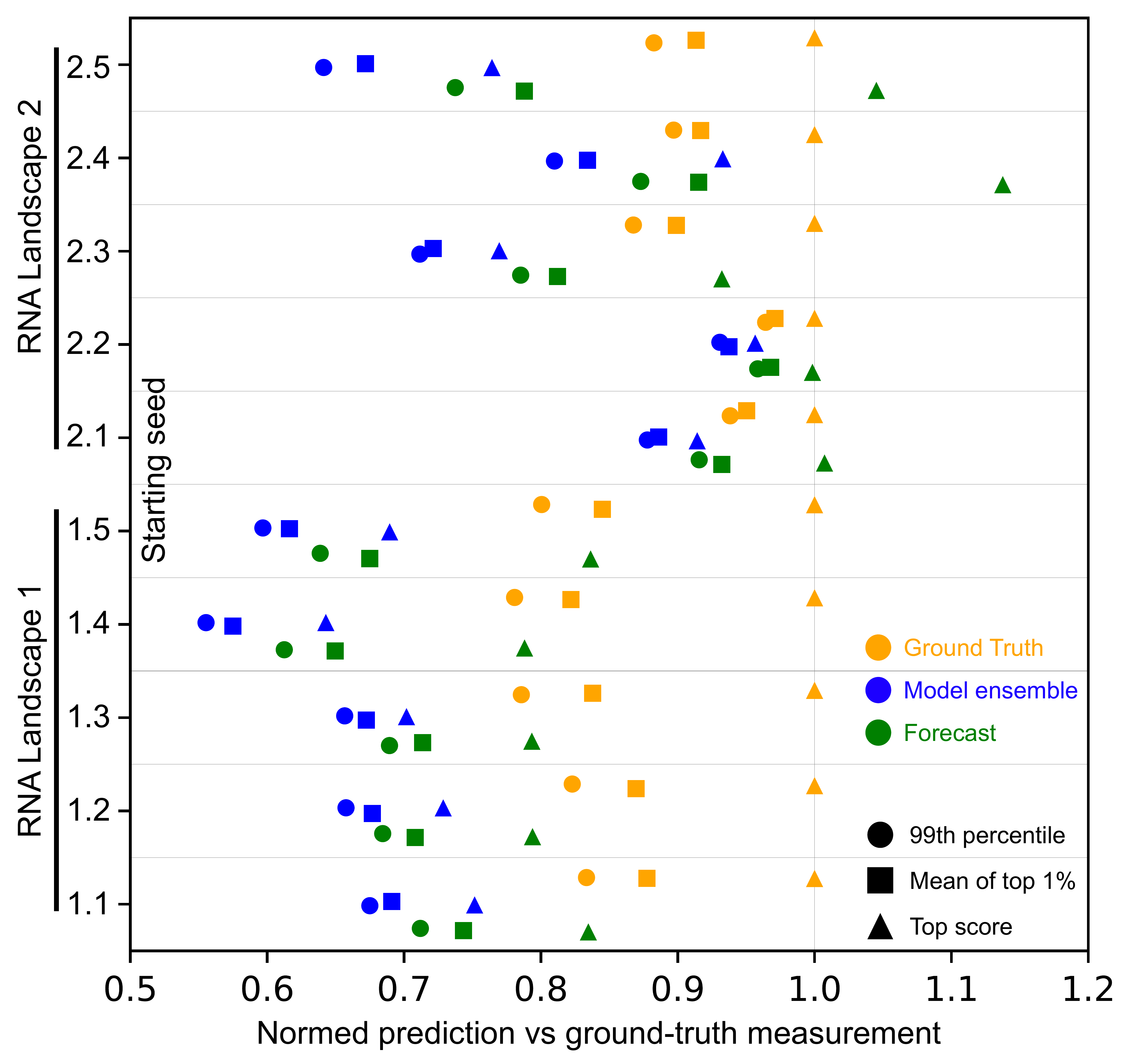}
  \caption{The ensemble, forecast, and true values for the 99th percentile, mean of the top percentile, and maximum value for the RNA experiments}
  \label{fig:rna_agg_plot}
\end{centering}
\end{figure*}

\subsection{Quantifying distribution shift}

As an attempt to quantify the difficulty of each forecasting problem, we computed metrics of covariate shift and label shift.  Covariate shift measures the change in distribution in covariate space (our sequences and covariates associated with those sequences), while label shift measures a change in the conditional distribution of the outcome given those covariates.  For this preliminary analysis, we restricted our study to using model ensemble scores as the main covariate of interest.  It would also be reasonable to apply this to edit distances, or higher-dimensional covarites.

To measure model score covariate shift, we can apply a 2-sample Kolmogorov-Smirnov test to the entire distributions of model scores for $\mathbf{S}^0$ and $\mathbf{S}^1$.  This gives us a measure on a common scale from no shift (0) to completely disjoint supports (1).

Measuring model score-based label shift precisely is challenging in our setting, since our data regularly violates the common assumption for label shift research that the test set output support is a subset of the training set support, so we cannot calculate ratios between the density functions.  Instead, we again us the 2-sample K-S test, this time comparing distributions of $\mathbf{Y}^0$ and $\mathbf{Y}^1$ but conditioned on high model scores (defined as the 90th percentile of the training set distribution and above).

\begin{table}
  \caption{Relative experimental difficulty due to model score-based covariate and label shift, as measured by the K-S score between distributions of training and test ensemble means, and between measurement distributions from among top-scoring variants, respectively}
  \label{tab:difficulty}
  \centering
  \begin{tabular}{p{0.2\linewidth}  p{0.2\linewidth} p{0.2\linewidth}} 
    \toprule
    Experiment     & Model score covariate shift   & Model score-based label shift \\
    \midrule
    AAV     & 0.332 & 0.159      \\
    GB1     & 0.960 & 0.916  \\
    GFP     & 0.888 & 0.541  \\
    \midrule
Landscape 1 Start 1 & 0.176 & 0.283  \\
Landscape 1 Start 2 & 0.162 & 0.227  \\
Landscape 1 Start 3 & 0.329 & 0.518  \\
Landscape 1 Start 4 & 0.579 & 0.575  \\
Landscape 1 Start 5 & 0.214 & 0.521  \\
Landscape 2 Start 1 & 0.371 & 0.606  \\
Landscape 2 Start 2 & 0.402 & 0.094  \\
Landscape 2 Start 3 & 0.283 & 0.830  \\
Landscape 2 Start 4 & 0.285 & 0.794  \\
Landscape 2 Start 5 & 0.381 & 0.747  \\
    \bottomrule
  \end{tabular}
\end{table}

We report these metrics in Table \ref{tab:difficulty}.We note that the AAV experiment, where the forecast performed especially well, had a lesser degree of covariate and label shift compared to other experiments.  At the other extreme, the GB1 experiment had extreme covariate and label shift, and while the forecasting method improved upon the ensemble prediction directionally, the forecast produced very low confidence interval coverage for this experiment.  This suggests a possible connection between shift scores and forecasting difficulty.  On the other hand, we can looking at the RNA experiments and consider one landscape at a time, which allows us to potentially isolate the relationship between these covariate shift metrics and forecasting performance.  Here, however, there does not appear to be any clear relationship between either type of distribution shift and the accuracy of the forecast.  Therefore, while the AAV and GB1 results suggests a possible connection, further experiments will be needed to validate these metrics as a useful tool for quantifying forecasting difficulty.

\subsection{Detailed forecasting algorithm}

See Algorithm \ref{alg:theta_inference} for a complete description of the forecasting algorithm described in Section \ref{sec:method-details} (excluding the extensions in Section \ref{sec:extensions}).

\begin{algorithm}[!htb]
   \caption{Inferring Gaussian mixture model parameters from a set of normally distributed model scores}
   \label{alg:theta_inference}
\begin{algorithmic}
   \STATE {\bfseries Input:} a training set ($\mathbf{S}^0$, $\mathbf{Y}^0$) and test set ($\mathbf{S}^1$) with model values $m_{ij}$ for each $s_i \in \mathbf{S}^0 \cup \mathbf{S}^1 $.
   \STATE {\bfseries Returns:} $(p_i, \mu_i^+, \mu_i^-, \sigma_i^+, \sigma_i^- )$ for each $i \in \mathbf{S}^1$
   \STATE Learn cutoff value $y_{mid}$ from $\mathbf{Y}^0$ using Otsu's method
   \FOR{$s_i$ {\bfseries in} $S^0$}
   \STATE Compute $\mu_i = \frac{\sum_j m_{ij}}{J}$ (model ensemble means)
   \STATE Compute $\sigma_i^2 = \frac{\sum_j (m_{ij}-\mu_i)^2}{J} $ (model ensemble variance)
   \STATE Compute $res_i^2 = (y_i-\mu_i)^2$ (squared residuals of model ensemble means)
   \STATE Compute  $I_i^+ = 1$ if $Y^0_i < y_{mid}$ and $0$ otherwise
   \ENDFOR
   \STATE Define $S^{0+} = \{i \mid Y^0_i \geq y_{mid} \}$ (training subset for ``functional'' sequences)
   \STATE Define $S^{0-} = \{i \mid Y^0_i < y_{mid} \}$ (training subset for ``broken'' sequences)
   \STATE Train isotonic model $IR_p$ on pairs $(\mu_i, I_i^+)$ for $i \in S^0$
   \STATE Train isotonic model $IR_{\mu^+}$ on  $(\mu_i, y_i)$ for $i \in S^{0+}$
   \STATE Train isotonic model $IR_{\mu^-}$ on  $(\mu_i, y_i)$ for $i \in S^{0-}$
   \STATE Train isotonic model $IR_{\sigma^+}$ on  $(\sigma_i^2, res_i^2)$ for $i \in S^{0+}$
   \STATE Train isotonic model $IR_{\sigma^-}$ on  $(\sigma_i^2, res_i^2)$ for $i \in S^{0-}$
   \FOR{$s_i$ {\bfseries in} $S^1$}
   \STATE Compute $\mu_i = \frac{\sum_j m_{ij}}{J}$ (model ensemble means)
   \STATE Compute $\sigma_i^2 = \frac{\sum_j (m_{ij}-\mu_i)^2}{J} $ (model ensemble variance)
   \STATE Compute $(p_i, \mu_i^+, \mu_i^-, \sigma_i^+, \sigma_i^- ) = (IR_p(\mu_i), IR_{\mu^+}(\mu_i), IR_{\mu^-}(\mu_i), IR_{\sigma^+}(\sigma_i), IR_{\sigma^-}(\sigma_i))$
   \ENDFOR
   \STATE {\bfseries return} $\{(p_i, \mu_i^+, \mu_i^-, \sigma_i^+, \sigma_i^- )\}$
\end{algorithmic}
\end{algorithm}

\end{document}